\documentclass[a4paper,aps,superscriptaddress,floatfix,nofootinbib,twocolumn]{revtex4-1}
\usepackage[utf8]{inputenc}
\usepackage{mathbbol}
\usepackage[pdftex]{graphicx}
\usepackage{latexsym,amsmath,verbatim,amssymb,txfonts}
\usepackage{color}
\usepackage{rotating}
\usepackage{verbatim}
\usepackage{multirow}
\usepackage[english]{babel}
\usepackage{comment}
\usepackage{dsfont}
\usepackage{bbold}
\usepackage{bbm}

\usepackage{graphicx}

\begin{document}

\title{Effective submodularity of influence maximization on temporal networks}
\author{\c{S}irag Erkol}
\affiliation{Center for Complex Networks and Systems Research, Luddy School of Informatics, Computing, and Engineering, Indiana University, Bloomington,
  Indiana 47408, USA}

\author{Dario Mazzilli}
\affiliation{Enrico Fermi Research Center, Via Panisperna 89 A, Rome, Italy.}
  
\author{Filippo Radicchi}
\affiliation{Center for Complex Networks and Systems Research, Luddy School of Informatics, Computing, and Engineering, Indiana University, Bloomington,
  Indiana 47408, USA}
\email{filiradi@indiana.edu}

\begin{abstract}
We study influence maximization on temporal networks. This is a special setting where the influence function is not submodular, and there is no optimality guarantee for solutions achieved via greedy optimization. 
We perform an exhaustive analysis on both real and synthetic networks. We show that the influence function of randomly sampled sets of seeds often violates the necessary conditions for submodularity. However, when sets of seeds are selected according to the greedy optimization strategy, the influence function behaves effectively as a submodular function. Specifically, violations of the necessary conditions for submodularity are never observed in real networks, and only rarely in synthetic ones. The direct comparison with exact solutions obtained via brute-force search indicate that the greedy strategy provides approximate solutions that are well within the optimality gap guaranteed for strictly submodular functions. Greedy optimization appears therefore an effective strategy for the maximization of influence on temporal networks.
\end{abstract}

\maketitle

\section{Introduction}

Influence maximization is the optimization problem aiming at finding the most influential nodes in a network~\cite{domingos2001mining, kempe2003maximizing}. The influence of a set of nodes is measured in terms of the size of the outbreak that the nodes generate when used as initial seeds for a spreading process occurring on the network.  Optimization is generally constrained by the number of initial spreaders. 

Influence maximization is a NP-hard problem~\cite{kempe2003maximizing}. The number of candidates to the solution increase exponentially with the number of seeds, so applying brute-force search to find the exact solution is possible for very small  networks only. Many attempts to find approximate solutions to the problem in a computationally feasible way are present in the literature.
Some approaches consist of using greedy algorithms~\cite{kempe2003maximizing, leskovec2007cost, chen2009efficient, nguyen2016stop, goyal2011celf, tang2015influence, minutoli2019fast}. 
Greedy optimization has provable performance bounds thanks to the fact that the outbreak size of a spreading process occurring on a static network is a submodular function~\cite{nemhauser1978analysis}. Roughly speaking, a set function is submodular if the addition one element to the input set generates an increment of the value returned by the function that decreases as the size of the input set increases. Unfortunately, greedy optimization algorithms are computationally expensive, thus
other more efficient approaches use centrality metrics to approximate the influence of nodes~\cite{erkol2019systematic, kitsak2010identification, de2014role, morone2015influence, lu2016vital, yang2020adaptive}. The drawback of centrality-based approaches is the lack of provable bounds of performance as they do not directly optimize the outbreak size of the spreading process.

The influence maximization problem is traditionally studied on static networks, i.e., the topology is given and does not change as spreading unfolds. However, there are many real-life networks where interactions between nodes might happen only for a given period of time~\cite{holme2012temporal}. If the timescales of the structural evolution and of spreading dynamics are similar, then changes in network topology can not be ignored as they can dramatically affect the outcome of the spreading process~\cite{prakash2010virus, karsai2011small, perra2012activity, valdano2015analytical}. So far, just a few paper considered the influence maximization problem on temporal networks~\cite{murata2018extended, han2017influence, zhuang2013influence, michalski2020effective}. Osawa \emph{et al.} studied the influence maximization problem on temporal networks for the susceptible-infected (SI) model \cite{osawa2015selecting}. They proposed an alternative to the greedy algorithm, and they showed that their method is very effective in correctly identifying top spreaders in networks with community structure. Michalski \emph{et al.} analyzed the effect of the size of time steps in temporal networks, including the aggregate version of the network~\cite{michalski2014seed}. They observed that using the aggregate version of the temporal network gives suboptimal results. Gayraud \emph{et al.} studied the independent cascade and the linear threshold models, showing that the
influence function is not submodular~\cite{gayraud2015diffusion}. They further showed that the activation time of nodes greatly affects the size of the final outbreak, and that delaying the activation of a node can increase its effective influence. 

In a recent paper, we studied the influence maximization problem under the susceptible-infected-recovered (SIR) model on temporal networks~\cite{erkol2020influence}. We performed a systematic analysis on $12$ real-world temporal networks, and analyzed the performances of different approximation methods that have different levels of knowledge on the network topology and dynamics. We found that complete knowledge of a network but in an aggregate way is not helpful in solving the problem effectively. On the other hand, knowledge of the initial stages of the network helps with finding good solutions to the influence maximization problem. The influence function of the SIR model on temporal networks is not submodular, except for the trivial case when the model is equivalent to the SI model.
However, the solutions provided by the greedy algorithm proved to be good upper bounds for the performances of other methods to identify influential spreaders. This fact suggests that, even though there is no theoretical proof on the performance of the greedy algorithm, in practice the optimization strategy is effective in approximating solutions to the influence maximization problem. This brings the questions of how often the condition for submodularity is violated, and how far the solutions of the greedy algorithm are from the ground-truth optimum. 

We answer the above questions in this paper. In particular,
we provide evidence for an effective submodular behavior of the influence function under greedy optimization.
Results of our analysis show that the condition for submodularity is violated frequently for randomly selected seeds; however, when seeds are selected with the greedy algorithm, the frequency drops to almost zero, especially for real-world networks. Also, we show that the solutions of the greedy algorithm have a performance very close to the optimal solution found with brute-force search.

\section{Methods}

\subsection{Temporal networks}

A temporal network is a collection of $T$ ordered network layers, namely $A(0), A(1), \ldots, A(t), \ldots, A(T-1)$, each representing the topology of the system at a specific time. All layers of a temporal network are composed of the same $N$ nodes, with labels uniquely identifying the nodes across the layers. 
The one above is just one possible definition of a temporal network; other definitions consider the evolution to happen in continuous rather than discrete time~\cite{holme2012temporal}.

There are $R(N, T) = 2^{{T\, {N \choose 2}}}$ total possible temporal networks with $N$ nodes and $T$ layers. There are in fact $2^{{N \choose 2}}$ different labeled networks with $N$ nodes; those networks can be permuted in $(2^{N \choose 2})^T$ ways, where permutations include also repetitions. 

We consider exhaustive enumerations of all possible temporal networks only in one experiment. Clearly, we choose very small values of the parameters $N$ and $T$ to make the enumeration computationally feasible. In all other experiments, the space of potential temporal networks is sampled by either constructing synthetic models or leveraging real data.

\subsubsection*{Synthetic temporal networks}

We generate random temporal networks with correlated layers. Specifically, the layer $A(t+1)$ of the temporal network is obtained by copying all the edges in the layer $A(t)$, and then shuffling with probability $r$ the end points of each individual edge with those of another random edge. For instance, if the edge $(i,j)$ undergoes shuffling, we first select at random another edge $(v,w)$. We then verify that  the edges $(i,w)$ and $(v,j)$ are not yet present in the network. If they are present, we select another edge $(v,w)$ and repeat the operation. Otherwise, we shuffle their end points in the sense that the edges $(i,j)$ and $(v,w)$ are removed from the network, and they are replaced by the edges $(i,w)$ and $(v,j)$. The above procedure of shuffling the end points of edges keeps the degree sequence of the network layers unchanged. For $r=0$, we have $A(t+1) = A(t)$ for all $t$, i.e.,  layers are perfectly correlated and the temporal network is essentially a static network. For $r=1$, all edges are surely shuffled, so no correlation exists between $A(t+1)$ and $A(t)$ except for the fact that they have the same degree sequence. In our experiments, we start by generating the first layer $A(0)$ according to the Erd\H{o}s-R\'enyi model with average degree $k$.

Also, we consider a simple uncorrelated model where all layers are generated according to the  Erd\H{o}s-R\'enyi model with average degree $k$. Networks of this type are statistically equivalent to those created according to model above with reshuffling probability $r=1$, with the only difference that, whereas the average degree of the network is invariant across network layers, the degree sequence is not preserved across network layers. In all cases where the value of the parameter $r$ is not specified, we will take advantage of this simple model to generate uncorrelated random temporal networks.

We note that the hypothesis of having an average degree that is invariant across layers is not necessarily a characteristic of real-world temporal networks~\cite{erkol2020influence}. To study the problem of influence maximization in realistic settings, we generate temporal networks directly from real data without the need of making any assumption, as described below.

\subsubsection*{Real-world temporal networks}
We use 12 empirical datasets containing time-stamped interactions between pairs of nodes. Some datasets contain only bidirectional interactions, while others consist of only unidirectional interactions. In both cases, we treat the interactions as undirected. To construct the temporal networks, we follow the exact same procedure followed in \cite{erkol2020influence}. Given the dataset, we divide the interactions into temporal windows of equal length. All interactions in a single slice are aggregated to create a temporal network layer that is undirected and unweighted. After the procedure is applied to each slice, we end up with $T$ layers of networks that form the temporal network as a whole. The networks used are listed in Table \ref{table:1}. For more details on the procedure of temporal network construction, see Ref.~\cite{erkol2020influence}.

\begin{table}[!htb]
\begin{center}
\begin{tabular}{|l|c|c|c|r|}\hline 
 Dataset & $T$ & $N$ & Ref. & Label \\\hline \hline 
Email, dept. 1 & $18$ & $309$ & \cite{paranjape2017motifs}  & a \\\hline
 Email, dept. 2 & $18$ & $162$ & \cite{paranjape2017motifs}& b \\\hline
 Email, dept. 3 & $18$ & $89$ & \cite{paranjape2017motifs} & c\\\hline
 Email, dept. 4 & $18$ & $142$ & \cite{paranjape2017motifs} & d\\\hline
 High school, 2011 & $11$ & $126$ & \cite{fournet2014contact} & e\\\hline
 High school, 2012 & $21$ & $180$ & \cite{fournet2014contact} & f\\\hline
 High school, 2013 & $14$ & $327$ & \cite{mastrandrea2015contact}& g \\\hline
 Hospital ward & $20$ & $75$ & \cite{vanhems2013estimating} & h\\\hline
 Hypertext, 2009 & $11$ & $113$ & \cite{isella2011s} & i\\\hline
 Primary school & $11$ & $242$ & \cite{gemmetto2014mitigation, stehle2011high}& j \\\hline
 Workplace & $20$ & $92$ & \cite{genois2015data} & k\\\hline
 Workplace-2 & $20$ & $217$ & \cite{genois2018can} & l\\\hline
\end{tabular}
\end{center}
\caption{{\bf Real-world temporal networks.} List of the empirical datasets used to construct temporal networks. From left to right, we report:  the name of the dataset, the number of network layers $T$, the number of nodes $N$ in the network, the reference to the paper(s) where the data were first considered, and the label of the dataset as used in the analysis of Fig.~\ref{fig:7}.}
\label{table:1}
\end{table}

\subsection{Spreading dynamics}

We consider the discrete version of the susceptible-infected-recovered (SIR) model for spreading dynamics. We use the model in the same way as already done in Ref.~\cite{erkol2020influence}. 
However, to properly compare the importance of the initial conditions on the long-term behavior of the model, we generate individual instances of the SIR model in a slightly different manner. Under this procedure, the difference in impact of different initial conditions is measured on identical, deterministic dynamical systems.  We then average the difference over multiple individual instances of the SIR model.

Before starting any dynamics on the network, we generate the $q$th instance of the SIR model with parameters $\lambda$ and $\mu$ by determining the propensity of individual edges to spread the infection and the propensity of individual nodes to recover at particular instants of time. Specifically, for each edge $(i,j)$ appearing at time $t$, we set the spreading propensity of the edge $\rho^{(q)}_{(i,j)}(t) =1$ with probability $\lambda$; otherwise, we set $\rho^{(q)}_{(i,j)}(t) =0$. Also, we set the recovery propensity of node $i$ at time $t$ as $\rho^{(q)}_{i}(t) =1$ with probability $\mu$, and $\rho^{(q)}_{i} (t)=0$ otherwise. 

Once propensities of edges and nodes are set for all temporal layers of the network, SIR dynamics can be run in a deterministic fashion starting from the initial condition $\vec{\sigma}(t=0) = [\sigma_1(t=0), \ldots, \sigma_N(t=0)]$, where $\sigma_i(0) = S, I$, or $R$. Please note that the initial condition does not depend on the actual realization of the SIR model. The rules that determine the dynamics for $t>0$ are as follows. Indicate with $\sigma_i^{(q)}(t)$ the dynamical state of node $i$ at time $t$ in the $q$th realization of the SIR model. We have that
\begin{equation}
    \sigma_{i}^{(q)}(t+1) = S \textrm{  if } \left\{
    \begin{array}{c} 
    \sigma_{i}^{(q)}(t) = S
    \\
    \bigwedge
    \\
     \nexists j \, | \, \sigma_{j}^{(q)}(t) = I \land  \rho_{(i,j)}^{(q)}(t) = 1 
    \end{array} \right. \; ,
    \label{eq:susc}
\end{equation}
meaning that node $i$ remains in the state $S$ if it does not get in active contact with any infected neighbor.
Also, we have that
\begin{equation}
\sigma_{i}^{(q)}(t+1) = I   \textrm{  if }  \sigma_{i}^{(q)}(t) = I \land \rho_{i}^{(q)}(t) = 0
\label{eq:infection1}
\end{equation}
and
 \begin{equation}   
    \sigma_{i}^{(q)}(t+1) = I   \textrm{  if } \left\{
    \begin{array}{c} 
    \sigma_{i}^{(q)}(t) = S \\
    \bigwedge
    \\
    \exists j \, | \,  \sigma_{j}^{(q)}(t) = I \land \rho_{i,j}^{(q)}(t) = 1
    \end{array} \right. \; .
    \label{eq:infection2}
\end{equation}
Eq.~(\ref{eq:infection1}) describes the case of a node already infected that does not recover. Eq.~(\ref{eq:infection2}) accounts instead for the change of the dynamical state of the node $i$ getting infected because of an active contact with at least one infected neighbor. Finally, we have
that 

\begin{equation}
    \sigma_{i}^{(q)}(t+1) = R  \textrm{  if } 
    \left\{
    \begin{array}{c} 
    \sigma_{i}^{(q)}(t) = I \land \rho_{i}^{(q)}(t) = 1 
    \\
    \bigvee
    \\
    \sigma_{i}^{(q)}(t) = R 
    \end{array} \right. 
    \; ,
\label{eq:recovery}
\end{equation}

i.e., node $i$ recovers if infected and prone to recovery at time $t$, or it does not change its state if it already recovered. 
After all the above operations are executed for all nodes $i$, time increases as $t \to t +1$.

Assuming that network evolution and spreading dynamics happen in discrete time and precisely at the same time scale simplify the numerical and analytical analysis of the dynamical system. We expect results obtained under these simplifications to be valid also for temporal networks evolving in continuous time as long as the duration of individual edges is sufficiently homogeneous and provided that the SIR model is reformulated in continuous time~\cite{pastor2015epidemic}. We recognize, however, that our framework should likely fail to properly describe SIR dynamics happening on temporal networks characterized by heterogeneous activity of the edges.

In our experiments, we restrict our attention to initial conditions where all nodes are in the susceptible state except of the nodes in the seed set $\mathcal X$ which are in the infected state, i.e.,  $\sigma_i(t=0) = I$ if $i \in \mathcal{X}$ and $\sigma_i(t=0) = S$ if $i \notin \mathcal{X}$. Starting from such initial configurations, the dynamics in a network with $T$ layers is simulated until the stage $T$. The outbreak size $f^{(q)}(\mathcal X)$ at the end of the $q$th realization of the process is calculated as the total number of infected and recovered nodes at $T$, i.e.,

$$ f^{(q)}(\mathcal X) = \frac{1}{N} \sum_{i=1}^N \left [ \mathds{1}_{\sigma_i^{(q)}(T),I} + \mathds{1}_{\sigma_i^{(q)}(T),R} \right ] \, ,$$
where $\mathds{1}_{x,y}$ is the identity operator, i.e., $\mathds{1}_{x,y}=1$ if $x=y$ and $\mathds{1}_{x,y}=0$ otherwise. 

We indicate the marginal gain in influence of adding  node $v$ to the seed set $\mathcal X$ as 
\begin{equation}
   f^{(q)}_{\mathcal X}(v) =  f^{(q)}(\mathcal X \cup \{v\}) - f^{(q)}(\mathcal X)\; .
    \label{eq:marginal_gain}
\end{equation}

We estimate the influence of the set $\mathcal X$ by taking the average 
value of the outbreak size over $Q$ independent realizations of the SIR model, i.e.,

\begin{equation}
    \left\langle f(\mathcal X) \right\rangle = \frac{1}{Q} \sum_{q=1}^Q f^{(q)}(\mathcal X) \; .
    \label{eq:influence}
\end{equation}

\subsection{Influence maximization}
Influence maximization is defined as the constrained optimization problem
\begin{equation}
    \mathcal{X}^*(M) = \arg \max_{|\mathcal{X}| = M} \left\langle f(\mathcal X)\right\rangle \; .
\label{eq:inf_max}
\end{equation}
Essentially, the goal of the problem is finding the set of seeds $\mathcal{X}^*(M)$ of size $M$ corresponding to the maximum value of the influence function.

Exact solutions of the problem are obtainable via brute-force search over all possible $\binom{N}{M}$ ways of choosing $M$ seed nodes out of $N$ total nodes in the network. For each of these candidate sets, the influence function of Eq.~(\ref{eq:influence}) should be evaluated. Clearly, the brute-force search can only be applied on relatively small networks and small seed set sizes. In most of the practical settings, the solution of the problem of Eq.~(\ref{eq:inf_max}) can be only approximated.

\subsection{Greedy optimization}
Approximate solutions to the problem can be obtained using a greedy optimization algorithm, defined as follows.
Representing the seed set at stage $M$ of the algorithm 
as $\mathcal G_M=\{ g_1,g_2,...,g_M\}$, we initialize 
$\mathcal G_0=\emptyset$. 
The seed at stage $M>0$ is chosen as
\begin{equation}
    g_M = \arg \max_{x \notin \mathcal G_{M-1}} \left\langle f_{\mathcal G_{M-1}} (x)\right\rangle \; .
    \label{eq:greedy}
\end{equation}
The algorithm of Eq.~(\ref{eq:greedy}) provides us with an approximation of the solution of Eq.~(\ref{eq:inf_max}) after $M=M$ iterations. As Eq.~(\ref{eq:greedy}) tells us, 
at each stage of the greedy selection algorithm, the node giving rise, on average, to the largest marginal gain in the influence function is selected. 
The average value of the marginal gain of each node $x \notin \mathcal G_{M-1}$ is numerically estimated from Eq.~(\ref{eq:influence}) by running independent simulations.

On static networks, the influence function of Eq.~(\ref{eq:influence}) is a  submodular function with non-negative marginal gains. These two properties guarantee that the solution provided by the greedy algorithm is at max $1-1/e$ times away from the ground-truth optimal solution~\cite{nemhauser1978analysis}. On temporal networks, such an optimality bound is guaranteed only for $\mu=0$. For $\mu>0$, the influence function is not 
necessarily a submodular function with non-negative marginal gains, hence there is no guarantee on the optimality gap for the solutions obtained via the greedy algorithm~\cite{gayraud2015diffusion}.

\section{Submodularity of the influence function}
In the following sections, we analyze the details for the violation of the condition for submodularity in temporal networks. In particular we show why the influence function is neither submodular nor $\gamma$-weakly submodular.

\subsection{Submodularity}
In their work, Kempe \emph{et al.} showed that the influence function of Eq.~(\ref{eq:influence}) on static networks has non-negative marginal gains and is a submodular function \cite{kempe2003maximizing}. 

The influence function has non-negative marginal gains if 
\begin{equation}
\label{eq:non-decreasing}
f_{\mathcal A}(v)  \geq 0 \; ,
\end{equation}
for any set $\mathcal A$ and for any node $v$. Influence is a submodular function if it satisfies the submodularity or ``diminishing returns'' condition, i.e., 
\begin{equation}
\label{eq:submodularity}
f_{\mathcal A}(v) \geq f_{\mathcal B}(v) \; ,
\end{equation}
for all nodes $v \notin \mathcal B$ and for all sets of nodes $\mathcal A\subseteq \mathcal B$. This means that the marginal gain obtained by adding node $v$ to the set $\mathcal A$, a subset of $\mathcal B$, must be greater than or equal to the marginal gain obtained by adding node $v$ to the set $\mathcal B$.

\begin{figure*}[!htb]
\centering
\includegraphics[width=\textwidth]{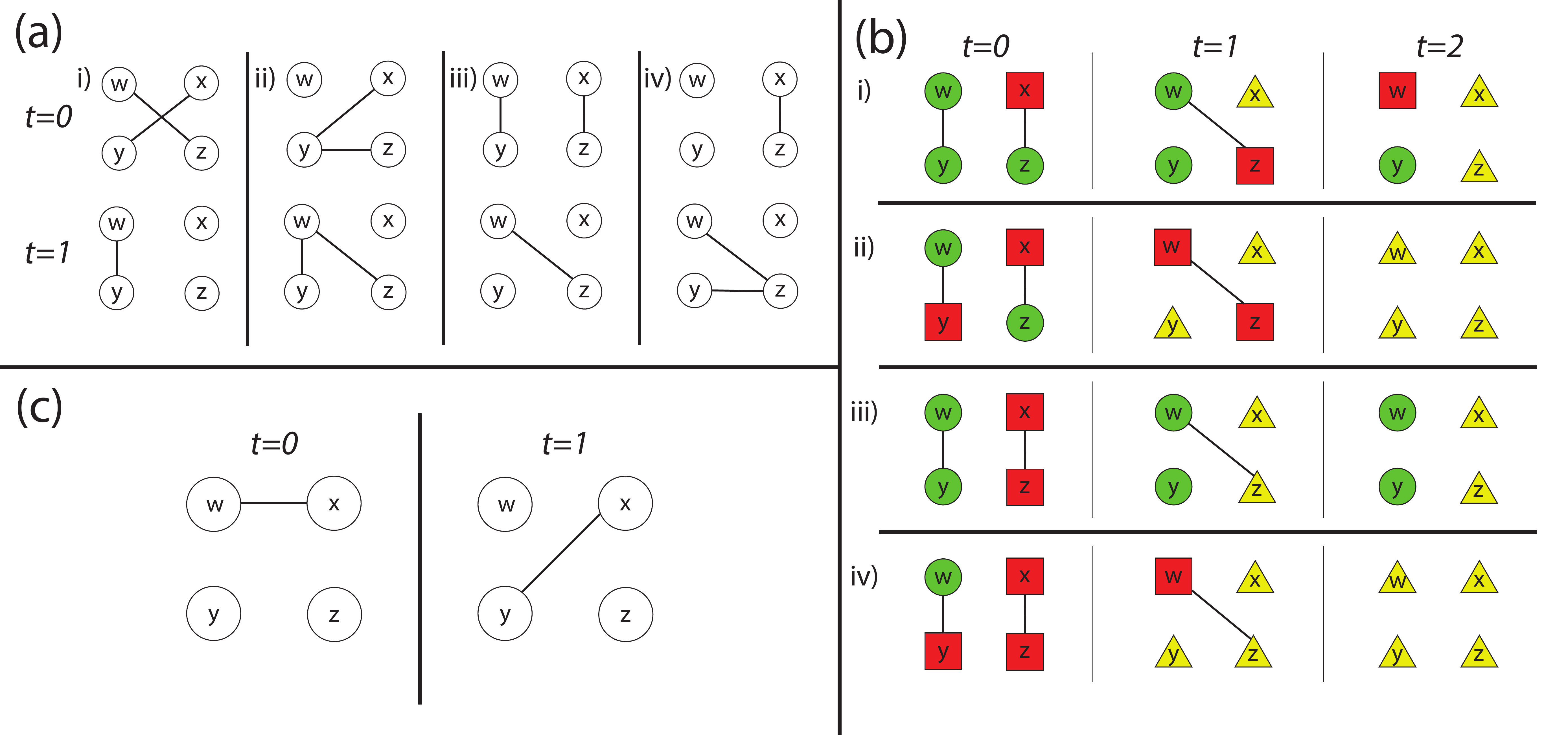}
\caption{
{\bf Violation of the necessary conditions for the
submodularity of the influence function on temporal networks.} For simplicity, we consider the deterministic case of the SIR model where the probabilities of infection and recovery are $\lambda=\mu=1$.
(a)  We display four possible scenarios for the marginal gain of adding node $v$ to sets $\mathcal A$ and $\mathcal B$, where $\mathcal A=\{x\}, \mathcal B=\{x,y\}, v=z$. The cases are separated on the basis of the marginal gains being negative or not. The marginal gains in the four scenarios are: (i) $f_\mathcal A(v)=1, f_\mathcal B(v)=2$, (ii) $f_\mathcal A(v)=1, f_\mathcal B(v)=-1$, (iii) $f_\mathcal A(v)=-1, f_\mathcal B(v)=0$, (iv) $f_\mathcal A(v)=-2, f_\mathcal B(v)=-1$. The inequality~(\ref{eq:submodularity}) is violated in (i), (iii), and (iv). (b) A counter-example for the submodularity of the influence function. Let $\mathcal A=\{x\}, \mathcal B=\{x,y\}, v=z$. Green circles denote nodes in the susceptible state, red squares denote nodes in the infected state, and yellow triangles denote nodes in the recovered state. (c) A counter-example for the $\gamma$-weakly submodularity of the influence function. Let $\mathcal A=\{w\}, \mathcal B=\{x,y,z\}$.}
\label{fig:1}
\end{figure*}

As we already mentioned, properties~(\ref{eq:non-decreasing}) and~(\ref{eq:submodularity}) hold for the SIR model on static networks. They hold for temporal networks too as long as the recovery probability is $\mu =0$. However, they are generally not valid on temporal networks for $\mu>0$. Violations of the condition of Eq.~(\ref{eq:submodularity}) may happen in three main ways, as illustrated in Fig.~\ref{fig:1}a. Four scenarios are realized depending on whether the marginal gains $f_\mathcal A(v)$ and $f_\mathcal B(v)$ are non-negative or negative. For $f_\mathcal A(v) \geq 0$ and $f_\mathcal B(v) < 0$ in (ii), the diminishing returns condition holds; for $f_\mathcal A(v) < 0$ and $f_\mathcal B(v) \geq 0$ in (iii), the diminishing returns condition is violated. In the other two scenarios  illustrated in panels (i) and (iv), it depends on the actual values of the marginal gain.

\begin{figure}[!htb]
\centering
\includegraphics[width=0.5\textwidth]{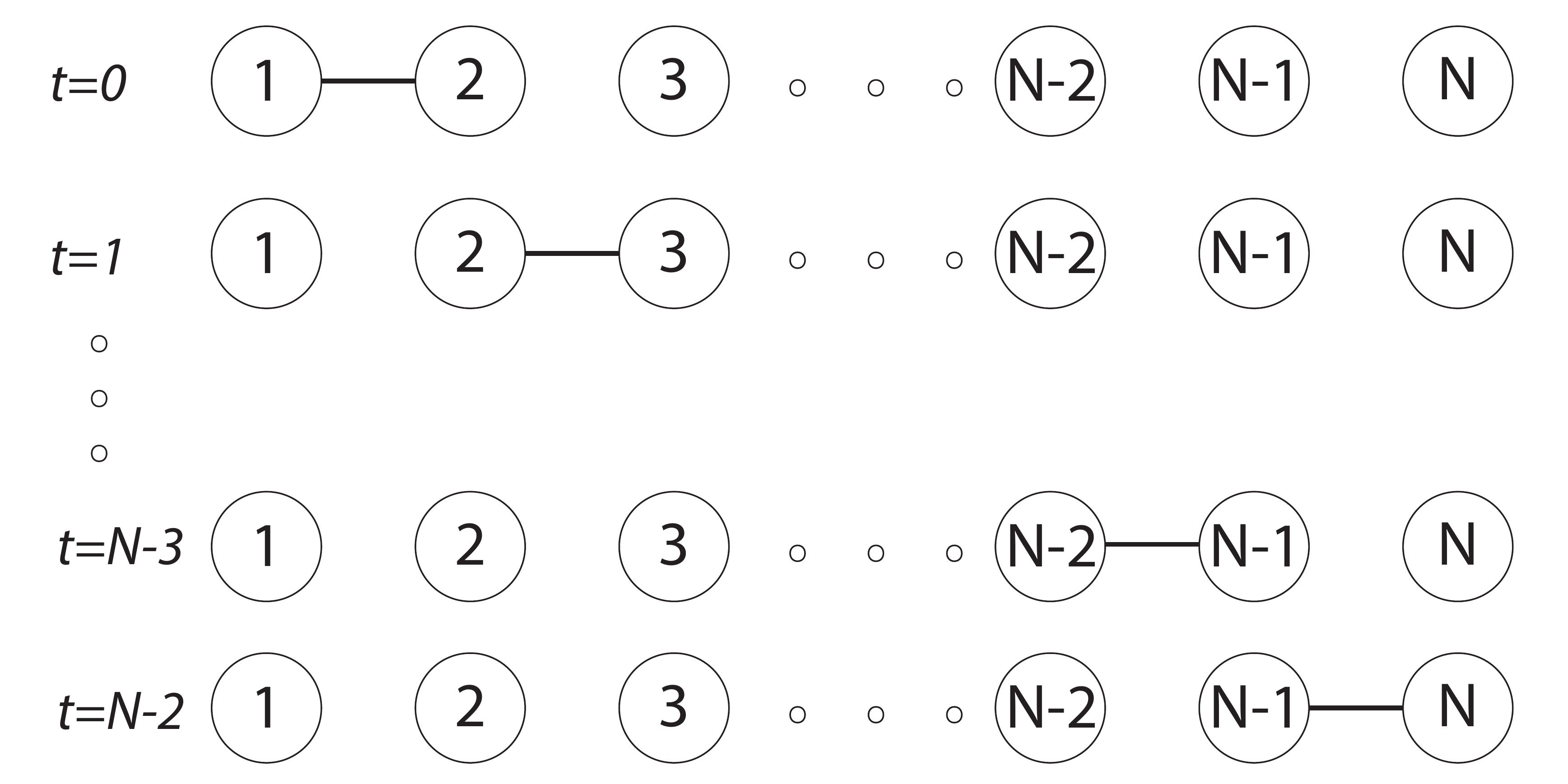}
\caption{
{\bf Blocking paths of future infections with multiple seeds.} We display a toy network where increasing the number of seeds have catastrophic effects on the outbreak size of the spreading process. For simplicity we consider the deterministic case of the SIR model where the probabilities of infection and recovery are $\lambda=\mu=1$. Setting node $1$ as the only seed of the process leads to maximum spread in the network, i.e., $f(\{1\}) = 1$. However, adding another node $i > 1$ to the seed set generates 
a reduction in the influence function, i.e., $f(\{1, i\}) = i/N$.}
\label{fig:2}
\end{figure}

The inspection of Fig.~\ref{fig:1}a reveals that violations of the conditions necessary for the submodularity property are caused by recovered nodes blocking the paths of future infections. In static networks, a recovered node would already have exploited any possible paths to infect its neighbors. The difference in temporal networks is that the neighbors of a node change in time. This means that the infection and recovery time of a node are fundamental factors that determine which paths are effectively used by the infection to propagate. In particular, an early infection of a node may be detrimental for the long-term fate of the spreading process just because once recovered the node may block paths that would have been otherwise available if the node was still in the susceptible state.

For example, in Fig.~\ref{fig:1}b we set $\mathcal A=\{x\}, \mathcal B=\{x,y\}$ and $v=z$. We also set $\lambda=\mu=1$ for simplicity. Each row shows the initial conditions at $t=0$ for the seed sets $\mathcal A, \mathcal A \cup v, \mathcal B$, and $\mathcal B \cup v$, respectively. Infections and recoveries occur at $t=0$ and $t=1$, and the final configuration is seen at $t=2$, where no more infections or recoveries happen. From Fig.~\ref{fig:1}b, we can see that $f(\mathcal A)=3$ in (i), $f(\mathcal B)=4$ in (ii), $f(\mathcal A \cup v)=2$ in (iii), and $f(\mathcal B \cup v)=4$ in (iv). Here, the condition of Eq.~(\ref{eq:submodularity}) is violated. The main reason of the violation is the premature infection of node $z$. The marginal gain of adding node $z$ to set $\mathcal A$ is $f_\mathcal A(z)=-1$. $z$  recovered at $t=1$, thus is not able to infect $w$. The infection could have occurred if the node was susceptible at the beginning of dynamics. So, by adding node $z$ to the set $\mathcal A$, the path to infecting node $w$ is blocked by the premature infection and recovery of node $z$, thus decreasing the total outbreak size at the end of the dynamics.

A toy example of how multiple, misplaced seeds may have dramatic consequences on the size of the outbreak is provided in Fig.~\ref{fig:2}. This is a rather specific and unrealistic example, where a single edge is present at each layer. We expect, however, that the very same issue, although in more complicated forms, is at the basis of violations of the necessary conditions for the submodularity of the influence function in real temporal networks.

\subsection{$\gamma$-weakly submodularity}
After showing that the influence function is not submodular, we can test weaker definitions of submodularity that would give us looser optimality gaps. In their work, Santiago and Yoshida defined $\gamma$-weakly submodularity for non-submodular functions \cite{santiago2020weakly}. In their definition, a function is $\gamma$-weakly submodular when

\begin{equation}
\label{eq:gammaweakly}
\sum_{v \in \mathcal B} f_\mathcal A(v) \geq \min \{ \gamma f_\mathcal A(\mathcal B), \frac{1}{\gamma} f_\mathcal A(\mathcal B) \}
\end{equation}
for any disjoint sets of nodes $\mathcal A$ and $\mathcal B$.
In the above inequality,
$0<\gamma \leq 1$. 
When the inequality holds, it is possible to find a solution with 
a so-called
randomized greedy algorithm 
resulting in
an optimality gap equal to $1-\gamma e^{-1/\gamma}$.

Unfortunately, the influence function of the SIR model on temporal networks is not $\gamma$-weakly submodular. An example of the violation of the condition of Eq.~(\ref{eq:gammaweakly}) is shown in Fig.~\ref{fig:1}c. We select $\mathcal A=\{w\}, \mathcal B=\{x,y,z\}$ and $\lambda=\mu=1$. The marginal gains are $f_\mathcal A(\mathcal B)=1, f_\mathcal A(x)=-1, f_\mathcal A(y)=0$, and $f_\mathcal A(z)=1$. The left hand side of the inequality~(\ref{eq:gammaweakly}) reads $\sum_{v \in \mathcal B} f_\mathcal A(v)=0$. This means that for the inequality to hold we need $\gamma=0$ or $\gamma=\infty$ because $f_\mathcal A(\mathcal B)=1$. However, the definition of  $\gamma$-weakly submodularity requires $0<\gamma \leq 1$, meaning that the inequality does not hold.

\section{Results}
There are temporal networks where the inequality~(\ref{eq:submodularity}) is violated, and the influence function is not submodular. 
In these situations, the greedy algorithm does not provide any guarantee on the optimality of its solutions. However, it is still possible that the algorithm provides solutions close enough to the ground-truth optimum. To investigate this property, we perform a systematic analysis on synthetic and real-world networks.

\subsection{Synthetic temporal networks}

\subsubsection{Random selection of seeds}

We calculate the frequency of violations of the 
inequality~(\ref{eq:submodularity}) as
\begin{equation}
\label{eq:submodviolation}
g(\mathcal A, \mathcal B, v) = \frac{1}{Q} \sum_{q=1}^Q 
H (f^{(q)}_\mathcal B(v) - f^{(q)}_\mathcal A(v)  ) \; ,
\end{equation}
where $\mathcal A \subset \mathcal B$ and $v \notin \mathcal B$,
and $H(x)$ is the Heaviside step function, i.e.,  $H(x)=1$ if $x>0$ and $H(x)=0$ otherwise.
Please note that $f^{(q)}_\mathcal A(v)$ and $f^{(q)}_\mathcal B(v)$ are both computed on the 
same $q$th instance of the SIR model.

\begin{figure}[!htb]
\centering
\includegraphics[width=0.45\textwidth]{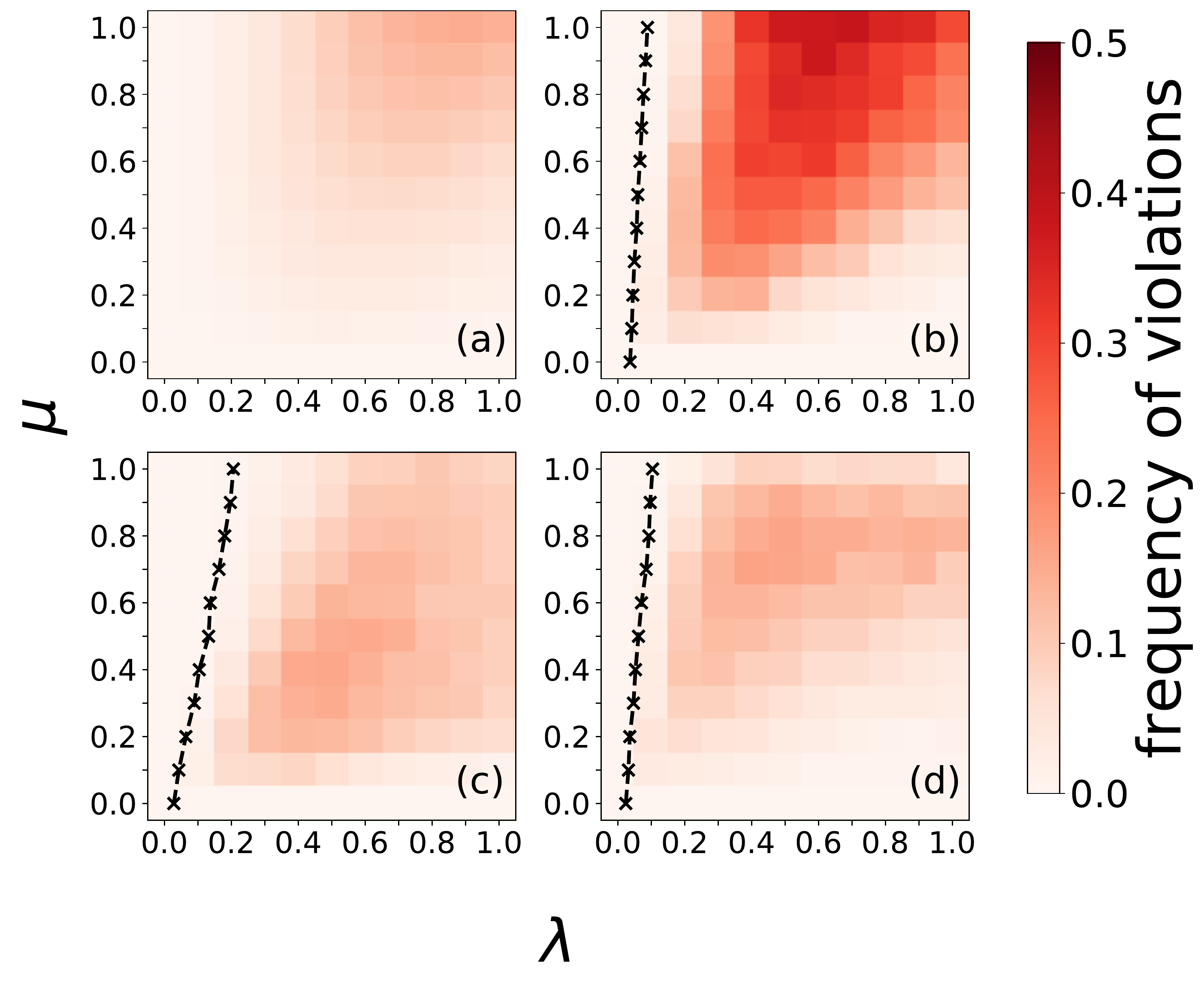}
\caption{{\bf Violations of the submodularity condition on temporal networks.}
(a) We display the frequency of violations of the 
diminishing returns inequality, i.e., $g$ as defined in Eq.~(\ref{eq:submodviolation}), on random synthetic networks of size $N=4$ as a function of the SIR parameters  
$\lambda$ and $\mu$. In the computation of Eq.~(\ref{eq:submodviolation}), the sets $\mathcal A$ and $\mathcal B$, and node $v$ are selected randomly with $|\mathcal A|=1, |\mathcal B|=2,$ and $\mathcal A \subset \mathcal B$. The simulations are run on a network with $N=4$ and $T=3$, and all possible configurations with this specific parameters have been used for the experiments. (b) Same as in panel (a), but only on a random temporal networks with $N=100, T=10,$ $k=5,$ and $r=1$. Results are averaged over 50 networks. The dashed black line shows the critical threshold values $\lambda_c(\mu)$ averaged over 10 networks. (c) Same as in panel (a), but for the real-world temporal network ``High school, 2012." The dashed black line shows the critical threshold values $\lambda_c(\mu)$. (d) Same as in panel (a), but for the real-world temporal network ``Hypertext, 2009."}
\label{fig:3}
\end{figure}

In Fig.~\ref{fig:3}a, we consider all the $R(N=4,T=3) = 262,144$ possible temporal networks  that can be formed with $N=4$ nodes and $T=3$ layers.
Indicating the four nodes of the network as $v, w, x$, and $y$,  we set $\mathcal{A}=\{y\}$ and $\mathcal{B} = \{x,y\}$. Please note that since we are considering all possible networks, the choice of the sets is irrelevant for our purposes. For a given network, we compute Eq.~(\ref{eq:submodviolation}) over $Q=100$ SIR model instances for each combination of $\mu$ and $\lambda$ values. We then take the average value of $g(\mathcal A, \mathcal B, v)$ over all possible networks to generate the heatmap of Fig.~\ref{fig:3}a. For $\lambda=0.00$, the influence function is submodular, there is no spreading, and $f_\mathcal A (v)=f_\mathcal B (v)=1$. For $\mu=0.00$, the spreading model becomes equivalent to the SI model, which displays a submodular influence function in temporal networks. For other $\lambda$ and $\mu$ values, violations of the inequality~(\ref{eq:submodularity}) are observed; maximum frequency of violations is registered for $\mu = 1.00$ and $\lambda = 0.90$.

We repeat a similar analysis on random temporal networks with $N=100, T=10, k=5,$ and $r=1$. Results are reported in Fig.~\ref{fig:3}b. For each run of the SIR model, we select three nodes at random, namely $x, y$, and $v$. We compose the sets $\mathcal{A}=\{y\}$ and $\mathcal{B} = \{x,y\}$, and compute Eq.~(\ref{eq:submodviolation}). Results in the figure are obtained by averaging $g(\mathcal A, \mathcal B, v)$ over $40$ runs of the SIR model, and over $50$ realizations of the random temporal network. The pattern revealed from the figure is similar to one obtained from the exhaustive analysis of Fig.~\ref{fig:3}a.

Finally, in Figs.~\ref{fig:3}c and \ref{fig:3}d, we report results for two real-world temporal networks. In this specific case, we still select nodes at random, namely $x, y$, and $v$ to compose the sets $\mathcal{A}=\{y\}$ and $\mathcal{B} = \{x,y\}$. We compute Eq.~(\ref{eq:submodviolation}) using $Q = 2,000$ SIR simulations. For each realization, we randomly select the nodes $x, y$, and $v$. The pattern observed is similar to those of the previous two cases, although we observe less violations than in the case of random temporal networks. Also, we observe that the  probability to observe a violation of the submodularity inequality is much higher in supercritical regime than in the subcritical regime. The dynamical regime of the SIR process on the network is supercritical if $\lambda > \lambda_c(\mu)$, and subcritical for $\lambda < \lambda_c(\mu)$. Here,  $\lambda_c(\mu)$ is the critical value of the spreading probability for a given value of the recovery probability $\mu$, see Appendix for details.

\begin{figure}[!htb]
\centering
\includegraphics[width=0.45\textwidth]{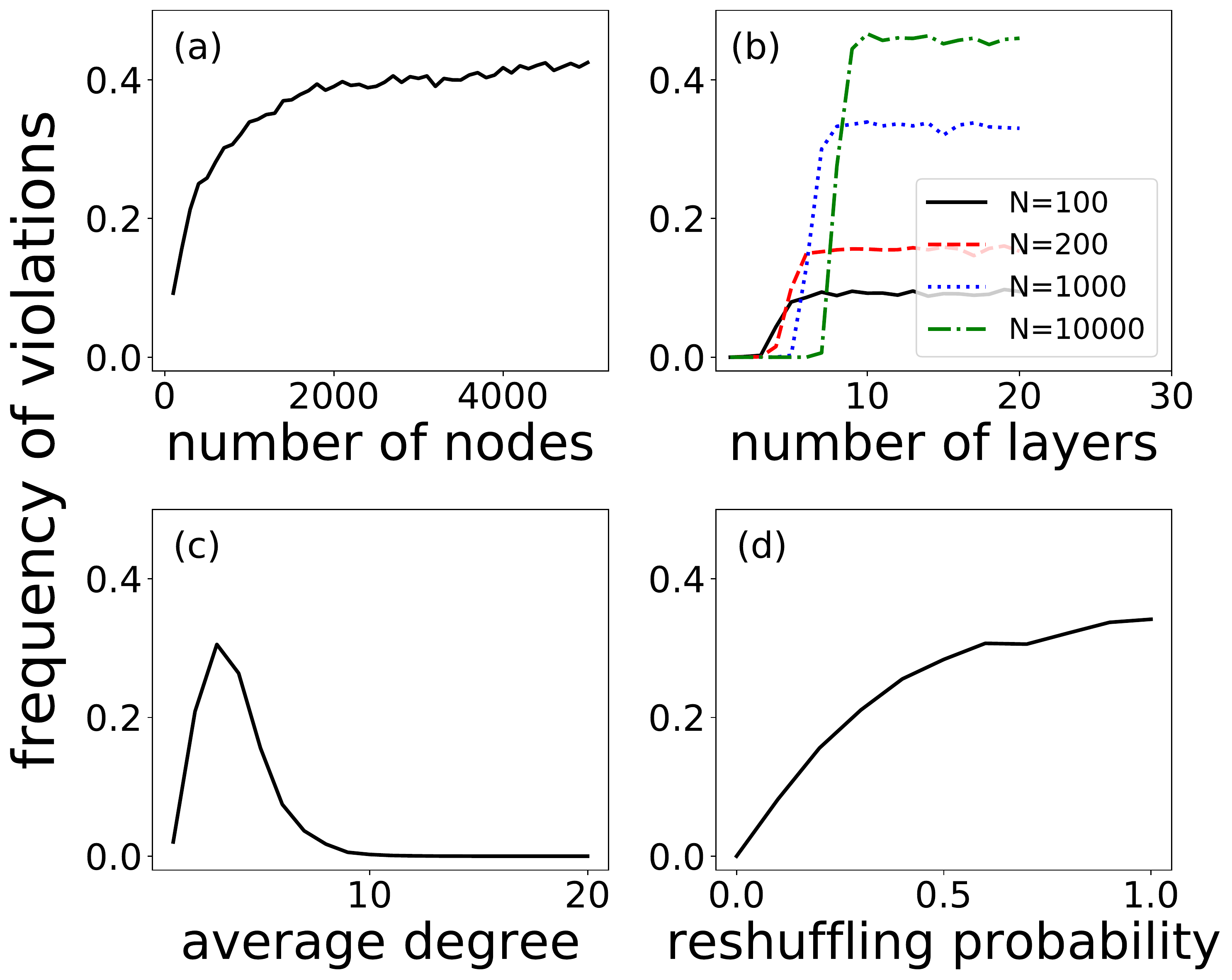}
\caption{{\bf Violations of the submodularity condition in synthetic temporal networks.} In all panels, unless stated otherwise, we consider $10,000$ networks composed of $N=200$ nodes, average degree $k=5$, total number of temporal layers $T=10$, and probability of edge shuffle between consecutive layers $r=0.2$. SIR parameters are $\lambda=\mu=1.00$. (a) We display $g$ in Eq.~(\ref{eq:submodviolation}) as a function of $N$. (b) We display Eq.~(\ref{eq:submodviolation}) as a function of $T$ for different values of $N$. (c) We display Eq.~(\ref{eq:submodviolation}) as a function of $k$. (d) We display Eq.~(\ref{eq:submodviolation})
as a function of $r$.}
\label{fig:4}
\end{figure}

We systematically study  violations of the submodularity condition in random temporal networks. In Fig.~\ref{fig:4}, we display the frequency of violations for different $N, T, k,$ and $r$ values. In each of our experiments, we first generate a network with a given set of parameters. Then, we select three nodes, namely $x$, $y$ and $v$, at random. We form the sets $\mathcal A = \{y\}$ and $\mathcal B = \{x, y\}$, and use these sets together with node $v$ in Eq.~(\ref{eq:submodviolation}) to tell whether the diminishing returns inequality is violated or not. Please note that we consider $\mu = \lambda =1.00$ in this set of experiments, so only one realization of the SIR model is possible. We consider $10,000$ networks, and record the average number of Eq.~(\ref{eq:submodviolation}) over such an ensemble. From Fig.~\ref{fig:4}, we see that as $N$ increases, the frequency of violation increases. For increasing $T$, we observe a phase transition from almost no violations to a non-null plateau value. For increasing $k$, there is a value where the frequency of violations of the diminishing returns property reaches a maximum; instead, when the network is either too sparse or too dense, the frequency drops to zero. As $r$ increases, we see an increase in the violation frequency. Note that for $r=0$, all the layers of the network are the same, meaning that the network is static and the influence function is submodular.

In Fig.~\ref{fig:4}, we reshuffle edges with probability  $r=0.2$. However, a similar phenomenology can be obtained by creating layers independently (see Fig.~\ref{fig:b1}). Given the similarity of the results, from now on, we focus our attention on the simpler model where random temporal networks are composed of layers generated independently.

Also, we measure the frequency of violations of the inequality~(\ref{eq:non-decreasing}) as

\begin{equation}
    \tilde g(\mathcal A, v) = 1 - \frac{1}{Q} \sum_{q=1}^Q H(f^{(q)}_\mathcal A(v)) \; .
    \label{eq:gainviolation}
\end{equation}
The above equation quantifies how often the addition of the node $v$ to the set $\mathcal A$ generates a marginal loss in the 
influence function. Results of our analysis are reported in Fig.~\ref{fig:b2}. We set $\lambda = \mu = 1.00$, and consider $Q = 1$ SIR simulations; in each simulation, we select at random two nodes, namely $x$ and $v$. We set $\mathcal{A} = \{x\}$. We estimate $\tilde g$ by taking the average over $10,000$ different networks; for each network we sample $\mathcal A$ and $v$ once.  We observe that the frequency for observing a marginal loss by adding a node follows a very similar behavior as the frequency of violations of the submodularity inequality, see Fig.~\ref{fig:4}. We take advantage of this similarity and focus our attention on marginal loss cases from now on, rather than measuring violations of the submodularity inequality.
This allows us to alleviate some computational burden without altering  the conclusions of the numerical analysis. We can in fact safely assume that the pattern of violations of the inequality~(\ref{eq:non-decreasing}) are similar to the pattern of violations of the condition of Eq.~(\ref{eq:submodularity}).

\begin{figure}[!htb]
\centering
\includegraphics[width=0.45\textwidth]{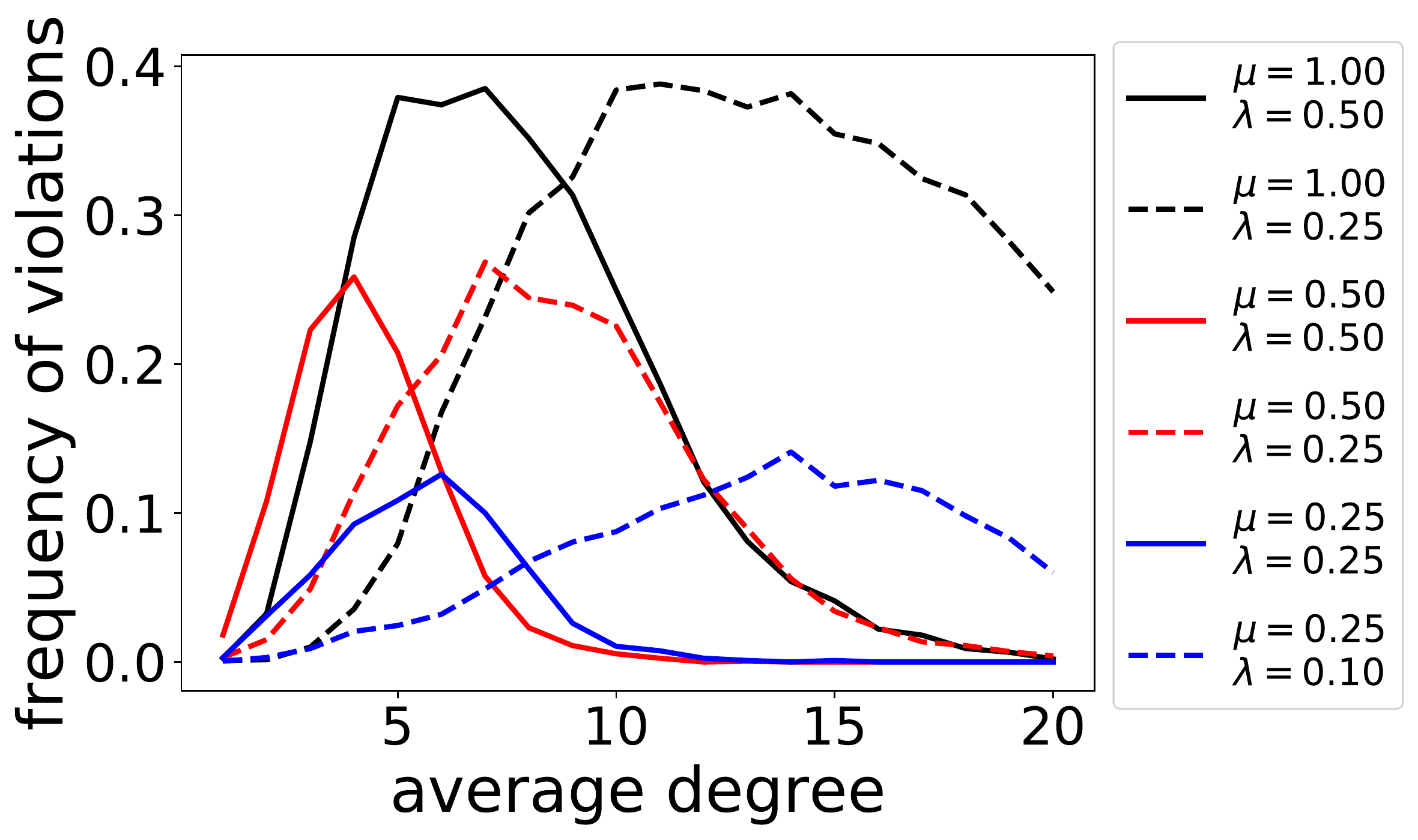}
\caption{{\bf Violations of the condition for marginal gain in synthetic temporal networks.} We measure the frequency of violations of the condition of marginal gain for the influence function using $\tilde g$ as defined in Eq.~(\ref{eq:gainviolation}). We analyze random temporal networks composed of $N=100$ nodes and $T=10$ layers. We consider different values of the average degree $k$, and of the SIR parameters $\lambda$ and $\mu$. In the estimation of Eq.~(\ref{eq:gainviolation}) we select $\mathcal A$ and $v$ randomly, and we average over $Q = 40$ instances of the SIR model. We further take the average over $50$ temporal networks. 
}
\label{fig:5}
\end{figure}

All the results presented until now used $\lambda=\mu=1.00$. In Fig.~\ref{fig:5}, we show 
results valid for $\lambda<1.00$ and $\mu \leq 1.00$. We see that the frequencies 
of inequality violations decrease as the recovery probability $\mu$
decreases, indicating that with a low recovery probability, it is less likely to observe 
violations of the inequality~(\ref{eq:non-decreasing}).

\subsubsection{Greedy selection of seeds}

\begin{figure}[!htb]
\centering
\includegraphics[width=0.45\textwidth]{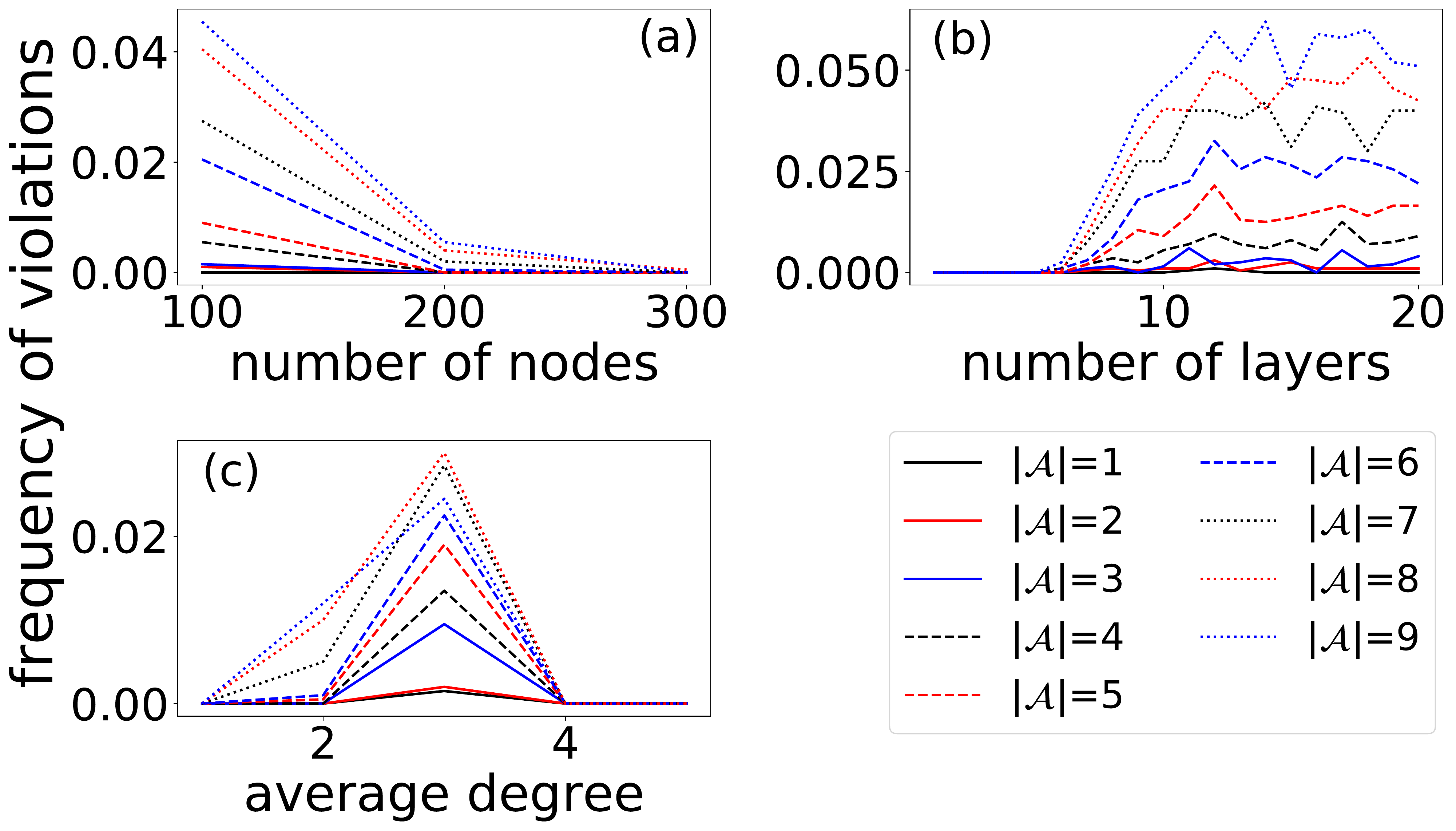}
\caption{{\bf Violations of the condition for marginal gain in synthetic temporal networks under greedy selection.} We measure the frequency of marginal losses using $\tilde g$ as defined in Eq.~(\ref{eq:gainviolation}) on random temporal networks for different sizes of $\mathcal A$. In all panels, unless stated otherwise, we consider networks composed of $N = 100$ nodes, $T = 10$ layers, and average degree $k = 2.5$. (a) We display Eq.~(\ref{eq:gainviolation}) as a function of $N$. (b) We display Eq.~(\ref{eq:gainviolation}) as a function of $T$. (c) We display Eq.~(\ref{eq:gainviolation}) as a function of $k$.}
\label{fig:6}
\end{figure}

All results we obtained so far indicate that violations of the inequalities~(\ref{eq:non-decreasing}) and~(\ref{eq:submodularity}) may occur frequently if tested for sets of randomly chosen nodes. It is, however, natural to ask whether such an observation is valid also when nodes are selected according to the greedy optimization protocol of Eq.~(\ref{eq:greedy}). In our tests, we measure the marginal gain obtained by adding the $M$th node at the $M$th stage of the greedy algorithm. We perform the tests on $2,000$ random temporal networks for various values of the parameters $N$, $T$, and $k$. SIR simulations are performed  for $\lambda = \mu = 1.00$. Results are obtained by averaging Eq.~(\ref{eq:gainviolation}) over the various network realizations, and are reported in Fig~\ref{fig:6}. The frequency of violations of the inequality~(\ref{eq:non-decreasing}) under greedy selection decreases significantly compared to the case where nodes are randomly selected. For the initial stages of the greedy algorithm, the frequency is almost zero; we only start seeing non-null cases of marginal loss in late stages of the algorithm. We do not show the results, but we observe that, for $\lambda<1.00$ and $\mu<1.00$, 
the frequency of marginal loss becomes almost zero when selecting up to 10\% of all nodes as initial spreaders.

\subsection{Real-world temporal networks}

\begin{figure}[!htb]
\centering
\includegraphics[width=0.45\textwidth]{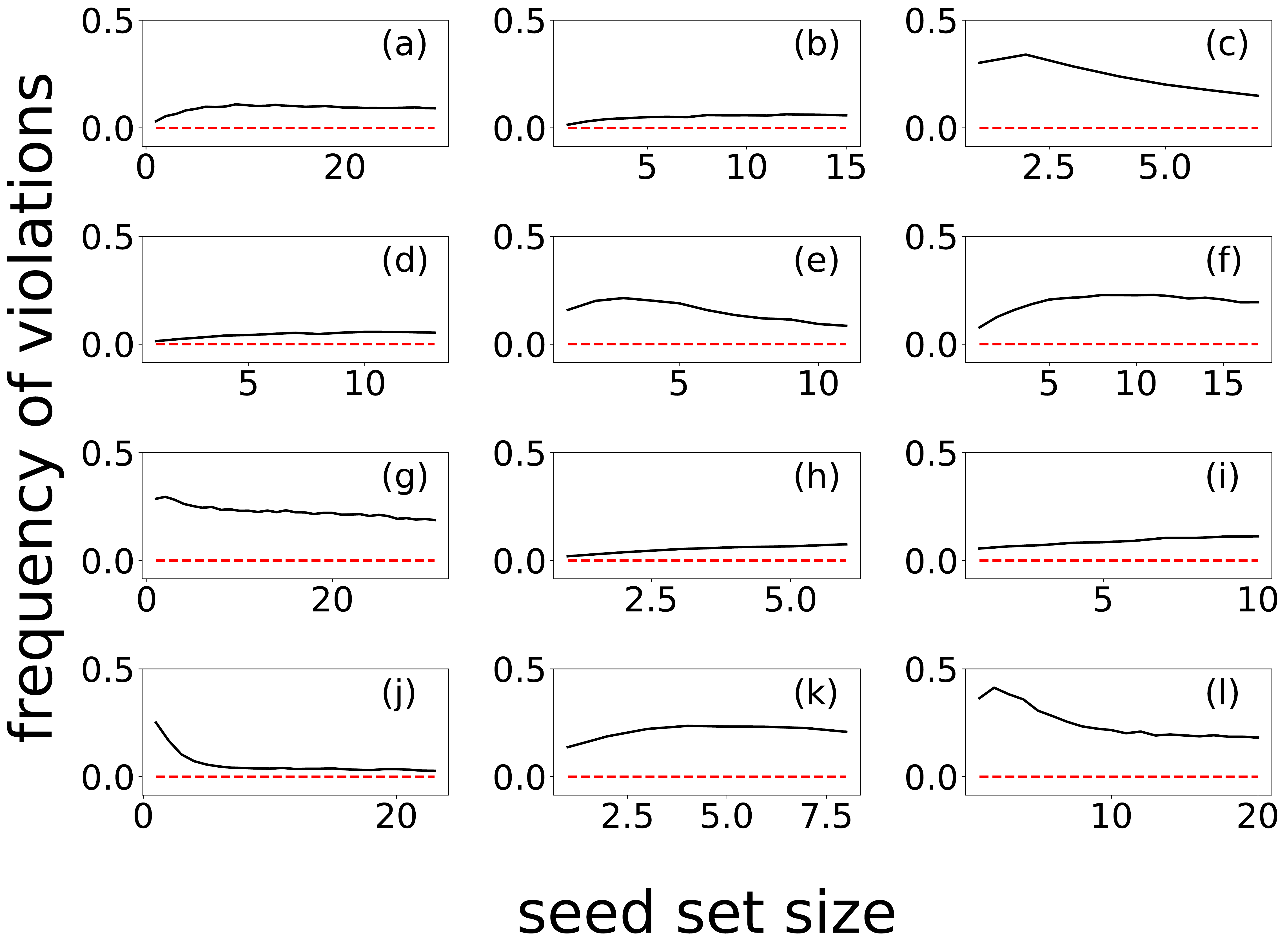}
\caption{
{\bf Violations of the condition for marginal gain in real temporal networks.} 
We measure the frequency of marginal losses using $\tilde g$ as defined in Eq.~(\ref{eq:gainviolation}) on
the real-world temporal networks listed in Table \ref{table:1}. Results are displayed as full black curves. Labels of the various panels reflect those appearing in the table. The SIR parameters are $\lambda = \mu = 1.00$, so that $Q=1$ in Eq.~(\ref{eq:gainviolation}). We select $\mathcal A$ and $v$ randomly $10,000$ times, and display the average value of the violation of marginal gains. The red dashed curves represent the frequency values when seeds are selected according to greedy optimization.}
\label{fig:7}
\end{figure}

We analyze the frequency of violations of condition of Eq.~(\ref{eq:non-decreasing}) on real temporal networks. We conduct our analysis on the networks shown in Table \ref{table:1}. Results of our analysis are reported in Fig.~\ref{fig:7}. We observe the same phenomenology as for the case of random temporal networks. If initial spreaders are selected randomly, then the number of configurations for which inequality~(\ref{eq:non-decreasing}) does not hold is not negligible. On the other hand, if spreaders are chosen according to the greedy strategy, then cases where the inequality~(\ref{eq:non-decreasing}) is violated are almost nonexistent. This finding suggests that, even though the right conditions to apply greedy optimization are not satisfied, in practice the greedy algorithm might still work as intended, finding solutions close to the ground-truth optimum.

\subsection{Greedy maximization against brute-force optimization}

\begin{figure}[!htb]
\centering
\includegraphics[width=0.45\textwidth]{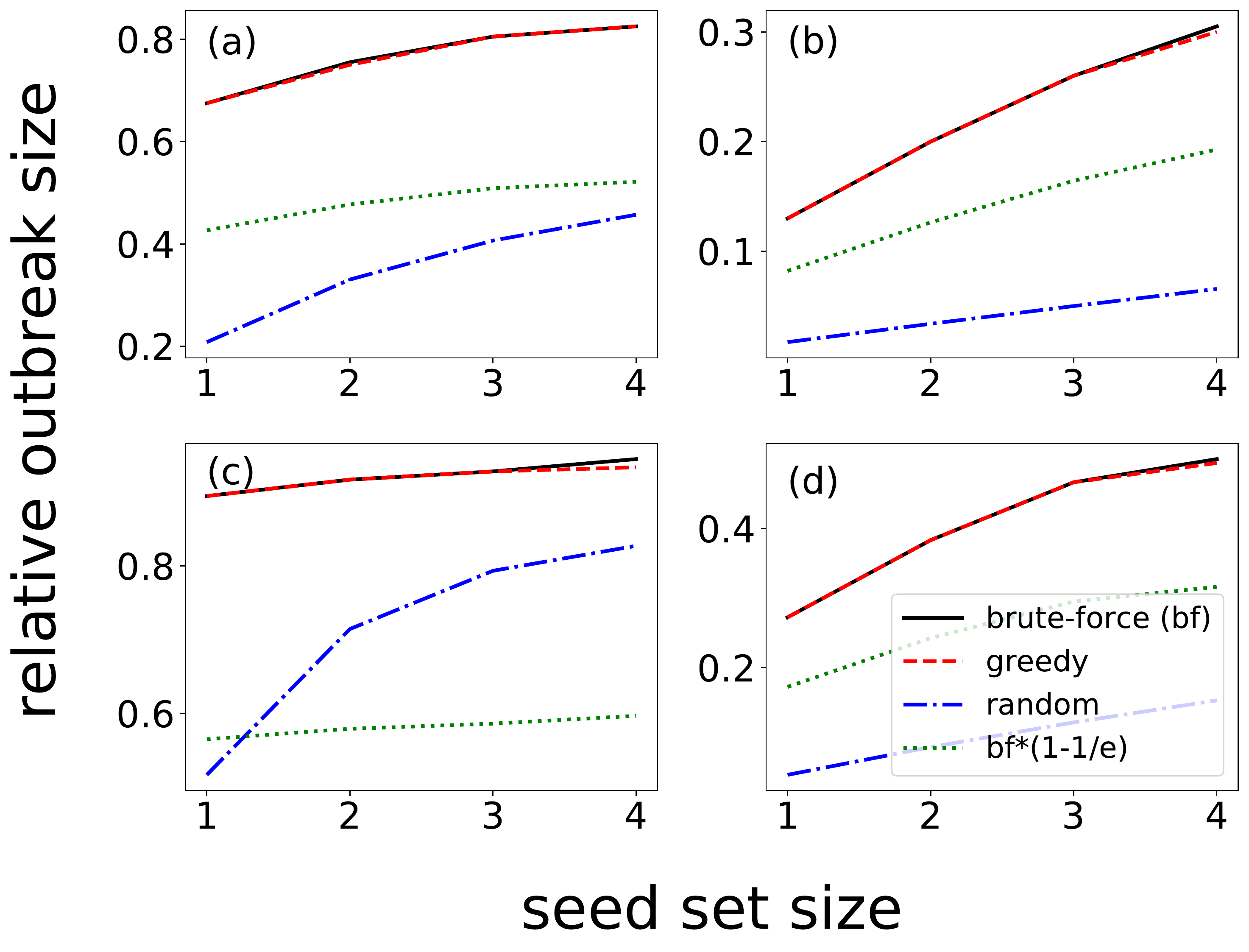}
\caption{
{\bf Optimal selection of influential spreaders in temporal networks.}
(a) We display the influence function of Eq.~(\ref{eq:influence}) as a function of the size of the set of influential spreaders $\mathcal X$. Different methods for the identification of the influential spreaders are used, either brute-force search (black), greedy optimization (red) or random selection (blue). We also display the $1-1/e$ bound from the brute-force solution (green). Results are valid for a random temporal network composed of
$N=200$ nodes, $T=10$ layers, and average degree $k=1.5$. SIR parameters are $\lambda=\mu=1.00$.
(b) Same as in (a) but for $\lambda=0.25$ and $\mu=0.50$. (c) Same as in (a) but for the real-world temporal network ``High school, 2012." (d) Same as in (c) but for $\lambda=0.10$ and $\mu=0.25$.
}
\label{fig:8}
\end{figure}

After showing that the greedy algorithm finds solutions characterized by negligible violations of the condition~(\ref{eq:non-decreasing}), we also want to see how close greedy solutions are to the ground-truth optimal solution. To this end we apply brute-force optimization to find seed sets of small size in both random and real temporal networks. We then compare these solutions to those obtained using greedy optimization and random selection. Results from all our tests are summarized in Tables~\ref{table:b1} and~\ref{table:b2}; results for a few sample cases are reported in Fig.~\ref{fig:8}.  In general, we observe that the greedy algorithm is almost optimal. In particular, the performance of the greedy algorithm is far above the $1-1/e$ bound that would be valid if the influence was indeed a submodular function with non-negative marginal gains. At the same time, we see that random selection performs poorly, generating outbreak sizes well below the optimality bound.

\section{Conclusion}
In spreading processes occurring on temporal networks, the influence function is not a submodular function, and yet  greedy optimization provides performances far better than those of other heuristic methods~\cite{erkol2020influence}. Here, we investigated the reasons behind such effectiveness of the greedy strategy. We measured violations of the necessary conditions for the submodularity property.
We observed that the premature infection and recovery of some nodes may have detrimental effects on the outbreak size, which in turn bring to violations of the necessary conditions for the submodularity property.
In our systematic analysis, we showed that violations occur frequently if seeds are selected at random. When seeds are selected according to the greedy optimization protocol, we observe that the frequency of violations is much lower in random temporal networks, and non-existent in real-world temporal networks. This finding suggests that, even though the function to be optimized does not satisfy the strict definition of a submodular function, in practice it behaves as an effectively submodular function under greedy optimization. As a matter of fact, solutions found by the greedy algorithm are as good as expected for optimization problems involving truly, mathematically speaking, submodular functions.
To actually test this hypothesis, we compared greedy solutions to ground-truth solutions in small networks and for small sizes of the seed sets. We observed that in all considered cases, the performance of the greedy algorithm is very close to the optimal solution. On average, the greedy algorithm has a performance of at worst $97\%$ in random temporal networks, and at worst $98\%$ in real-world temporal networks. This is much higher than the optimality guarantee of $63\%$ for greedy algorithm in static networks and also much better than the performance that can be achieved with randomly selected nodes. 
The fact that greedy optimization generates quasi-optimal solutions to the influence maximization problem on temporal networks is associated with its ability of avoiding the selection of seeds whose premature recovery would block future infection paths. We believe that this ability is not related to the specific protocol of optimization, rather to the fact that the optimization procedure relies on direct measurements of the influence function. We expect that any other optimization algorithm that uses dynamical information should able to achieve similar performance by learning about the blocking effect of some nodes
 via measurements of the influence function. Also, 
the great effectiveness displayed by the greedy optimization strategy might indicate the effective existence of a tight lower bound on the performance of the greedy algorithm in solving the influence maximization problem on temporal networks. We leave to future research the challenging task to formulate a theory for such a performance guarantee.

\acknowledgments{
\c{S}.E. and F.R. acknowledge support by the Air Force Office of Scientific Research (FA9550-21-1-0446); F.R.  acknowledges support by the Army Research Office (W911NF-21-1-0194). The funders had no role in study design, data collection and analysis, decision to publish, or  any opinions, findings, and conclusions or recommendations expressed in the manuscript.}

\appendix

\section{Finding the critical threshold}
\renewcommand{\thefigure}{A\arabic{figure}}
\setcounter{figure}{0}
\renewcommand{\thetable}{A\arabic{table}}
\setcounter{table}{0}

The critical threshold value $\lambda_c$ of a temporal network is a function of the recovery probability $\mu$, denoted as $\lambda_c(\mu)$. In order to estimate $\lambda_c(\mu)$, we start the spreading process from each single node in the first layer of the temporal network, and calculate the outbreak size. We repeat this process for 500 times for each node for a range of values for the infection probability $\lambda$. The $\lambda$ value that gives the maximum for the ratio between standard deviation and mean of outbreak sizes is selected as the critical threshold value $\lambda_c$ for a given recovery probability $\mu$.

\section{Additional Results}
\renewcommand{\thefigure}{B\arabic{figure}}
\setcounter{figure}{0}
\renewcommand{\thetable}{B\arabic{table}}
\setcounter{table}{0}

\begin{figure}[!htb]
\centering
\includegraphics[width=0.45\textwidth]{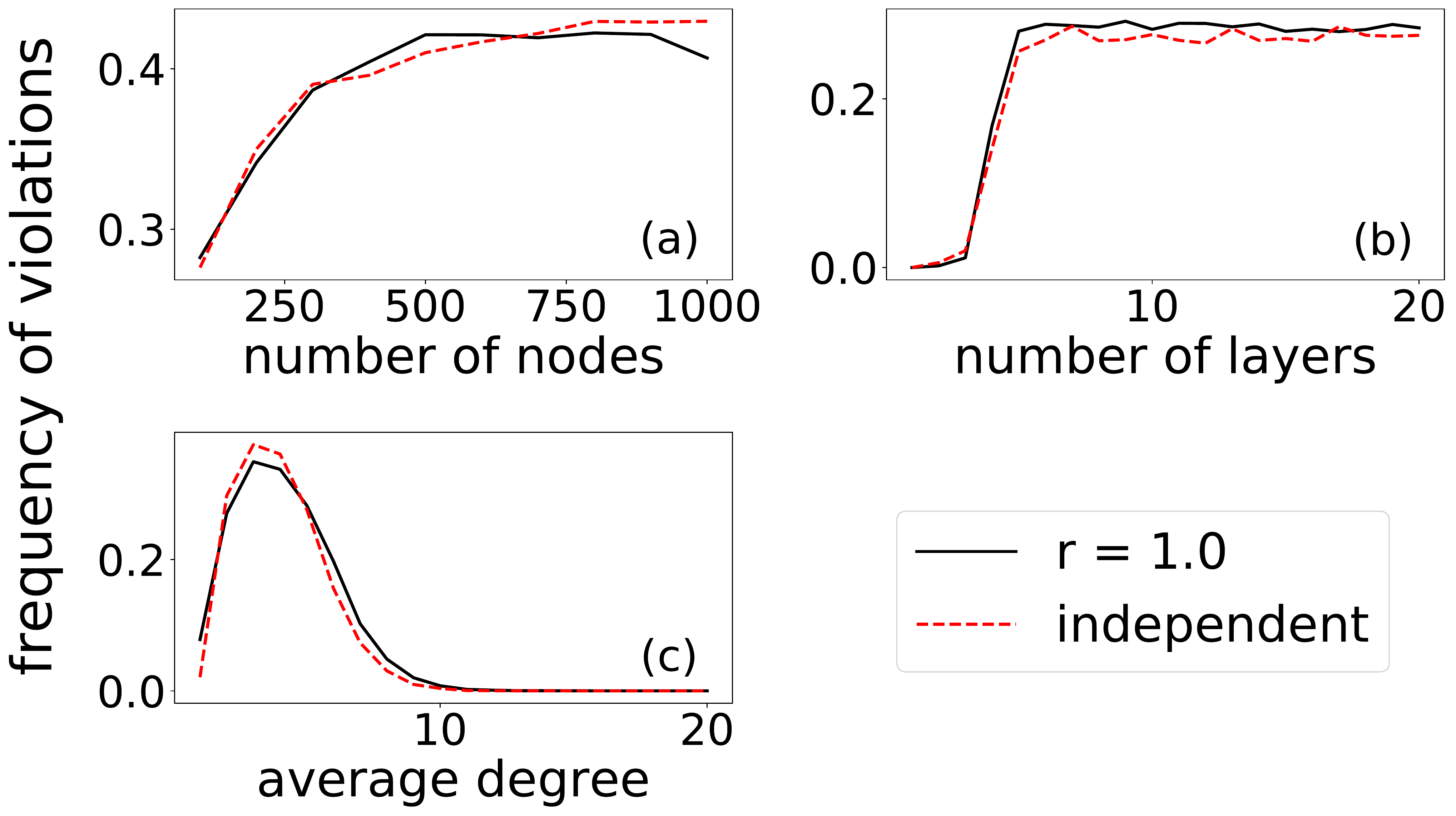}
\caption{{\bf Violations of the submodularity condition on synthetic
temporal networks.} Frequency of violations of the inequality~(\ref{eq:submodularity}) on synthetic network models with $N=100, T=10, k=5$ unless stated otherwise, and SIR parameters $\lambda=\mu=1.00$. (a) Eq.~(\ref{eq:submodviolation}) as a function of $N$. Results obtained for $r=1.0$ (black curve) are compared to the results obtained when layers in the temporal network are created independently (red curve). (b) Same as in (a) but we display Eq.~(\ref{eq:submodviolation}) as a function of $T$. (c) Same as in (a) but for Eq.~(\ref{eq:submodviolation}) as a function of $k$.}
\label{fig:b1}
\end{figure}

\begin{figure}[!htb]
\centering
\includegraphics[width=0.45\textwidth]{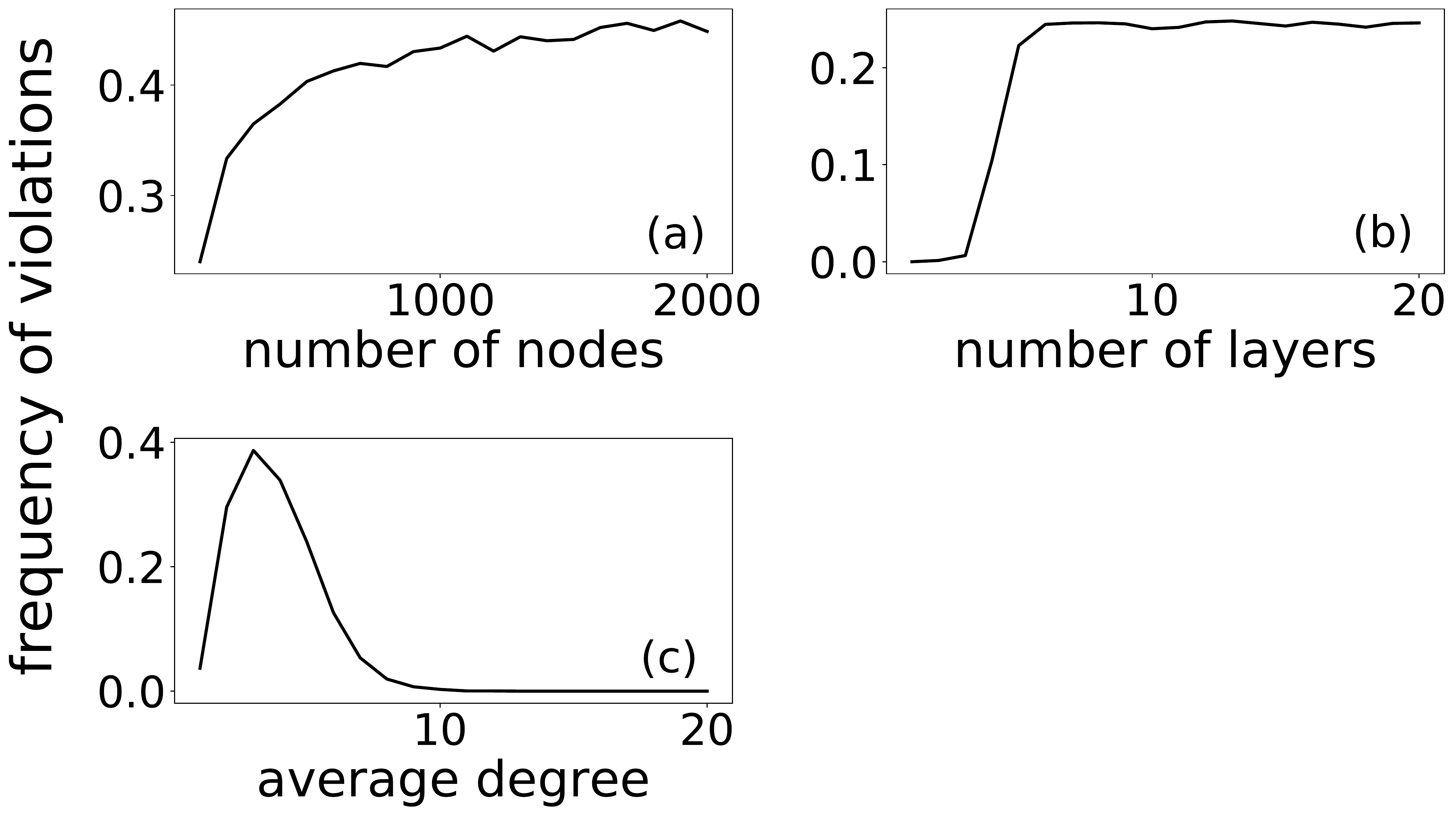}
\caption{{\bf Violations of the condition for marginal gain in synthetic temporal networks.} Frequency of marginal loss cases with random seeds on random temporal networks for $N=100, T=10, k=5$ unless stated otherwise, and SIR parameters $\lambda=\mu=1.00$. (a) Eq.~(\ref{eq:gainviolation}) as a function of $N$. (b) Eq.~(\ref{eq:gainviolation}) as a function of $T$. (c) Eq.~(\ref{eq:gainviolation}) as a function of $k$.}
\label{fig:b2}
\end{figure}

In Fig.~\ref{fig:b1}, we report the frequency of  violations of the inequality~(\ref{eq:submodularity}) on synthetic network models for different values of the parameters $N,T,$ and $k$. The spreading process is initiated by randomly selected seeds. Specifically, we compare the results for a temporal network created with $r=0.2$ against a network created with independent layers. In Fig.~\ref{fig:b2}, we show the results for the frequency of violations of the inequality~(\ref{eq:non-decreasing}) on synthetic network models for different values of the parameters $N,T,$ and $k$. Also here, the spreading process is initiated by randomly selected seeds. In Tab.~\ref{table:b1}, we report the average performance of greedy algorithm relative to brute-force optimization in real-world temporal networks. In Tab.~\ref{table:b2}, we report the same results for random temporal networks.

\begin{table}[!htb]
\begin{center}
\begin{tabular}{|r|r|c|c|c|c|}\hline
$\lambda$ & $\mu$ & $|\mathcal A|=2$ & $|\mathcal A|=3$ & $|\mathcal A|=4$ & $|\mathcal A|=5$ \\\hline
$1.00$ & $1.00$  & 99.0\% & 98.8\% & 98.4\% & 98.2\% \\\hline
$0.50$ & $1.00$  & 99.6\% & 99.3\% & 99.1\% & 98.7\% \\\hline
$0.25$ & $1.00$  & 100.0\% & 99.3\% & 98.9\% & 98.5\% \\\hline
$0.50$ & $0.50$  & 99.7\% & 99.6\% & 99.5\% & 99.4\% \\\hline
$0.25$ & $0.50$  & 99.6\% & 99.6\% & 98.9\% & 98.8\% \\\hline
$0.10$ & $0.50$  & 98.8\% & 98.3\% & 98.6\% & 98.5\% \\\hline
$0.25$ & $0.25$  & 98.9\% & 99.3\% & 99.1\% & 99.1\% \\\hline
$0.10$ & $0.25$  & 99.6\% & 99.1\% & 99.3\% & 99.3\% \\\hline
$0.05$ & $0.25$  & 99.2\% & 99.0\% & 98.6\% & 98.7\% \\\hline
\end{tabular}
\end{center}
\caption{{\bf Greedy selection of optimal spreaders in real temporal networks.} We report the average outbreak size of the seeds found by greedy algorithm on various networks relative to the optimal solution found with brute-force optimization. The results are averaged over all real-world temporal networks listed in Tab.~\ref{table:1}. We excluded ``Email, dept. 1" and ``High school, 2013" due to their size.}
\label{table:b1}
\end{table}

\begin{table}[!htb]
\begin{center}
\begin{tabular}{|r|r|c|c|c|}\hline
$\lambda$& $ \mu$ & $|\mathcal A|=2$ & $|\mathcal A|=3$ & $|\mathcal A|=4$ \\\hline
$1.00$& $ 1.00$  & 99.2\% & 98.4\% & 98.1\% \\\hline
$0.50$& $ 1.00$  & 97.9\% & 97.6\% & 97.1\% \\\hline
$0.25$& $ 1.00$  & 100.0\% & 100.0\% & 100.0\% \\\hline
$0.50$& $ 0.50$  & 99.6\% & 99.4\% & 98.7\% \\\hline
$0.25$& $ 0.50$  & 99.7\% & 99.3\% & 98.9\% \\\hline
$0.10$& $ 0.50$  & 100.0\% & 100.0\% & 100.0\% \\\hline
$0.25$& $ 0.25$  & 99.4\% & 99.4\% & 99.4\% \\\hline
$0.10$& $ 0.25$  & 99.7\% & 99.8\% & 99.8\% \\\hline
$0.050$& $ 0.25$  & 100.0\% & 100.0\% & 100.0\% \\\hline
\end{tabular}
\end{center}
\caption{{\bf Greedy selection of optimal spreaders in synthetic temporal networks.} We report the average outbreak size of the seeds found by greedy algorithm on random temporal networks relative to the optimal solution found with brute-force optimization. The results are averaged over random temporal networks created with all combinations of the parameters $N=200, T \in \{5,10\},$ and $k \in \{1.3,1.4,1.5,1.6,1.7,2.0,2.5\}.$}
\label{table:b2}
\end{table}

\clearpage

\bibliographystyle{ieeetr}
\bibliography{references}

\end{document}